\documentclass[amsmath,amssymb,reprint,bibnotes]{revtex4-2}
\usepackage{graphicx}
\usepackage{bm}
\usepackage{amsmath} 
\usepackage{graphics}
\usepackage[dvipsnames]{xcolor}
\usepackage[mathlines]{lineno}
\usepackage{braket}
\usepackage[compact]{titlesec}
\begin{document}
\title{Optically tunable linear and nonlinear enhancement of index of refraction}
\author{Elif Ozturk$^{\bf (1,2) }$}
\author{Hira Asif$^{\bf (3,4) }$}
\author{Mehmet Gunay$^{\bf (5)}$}
\author{Mehmet Emre Tasgin$^{\bf (1)}$}
\author{Ramazan Sahin$^{\bf (3,4)}$}\email{rsahin@itu.edu.tr}

\affiliation{${\bf (1)}$ {Institute of Nuclear Sciences, Hacettepe University, 06800 Ankara, Turkey}}
\affiliation{${\bf (2)}$ {Department of Nanotechnology and Nanomedicine, Graduate School of Science and Engineering, Hacettepe University, 06800 Ankara, Turkey}}
\affiliation{${\bf (3)}$ {Department of Physics, Akdeniz University, 07058 Antalya, Turkey}}
\affiliation{${\bf (4)}$ {Türkiye National Observatories, TUG, 07058 Antalya, Turkey}}
\affiliation{${\bf (5)}$ {Department of Nanoscience and Nanotechnology, Faculty of Arts and Science, Burdur Mehmet Akif Ersoy University,
		Burdur 15030, Turkey}}

\date{\today}

\begin{abstract}
Control of optical properties of materials by tuning their refractive index can revolutionize the current state-of-the-art technology to manipulate light propagation in the high loss media. Here we demonstrate active optical tuning of the plasmonic analog of \textit{enhancement of index of refraction} (EIR) in both linear and nonlinear regimes using a quantum mechanical approach. By employing a pump-probe scheme, we investigate the tuning of refractive index of the probe field by varying amplitude and phase of the pump source. In contrast to classical approach used in \cite{Panahpour2019}, we formulate both first- and second-order quantization to analyze nonlinear enhancement in the refractive index by modulating the response function of probe field. This approach enables indirect tuning of nonlinear modes and coherent control of the probe pulse under the coupling of linear plasmonic modes supported by two L-shaped nano-ellipsoids. Varying the pump amplitude not only shows a significant enhancement in the EIR in both regimes but also effectively suppresses optical losses with zero dispersion at the system's resonance frequency. Additionally, tuning pump phase induces a spectral shift in the frequency of the probe field which open new ways for active tuning of epsilon-near-zero (ENZ) materials. Our approach offers all-optical tuning of nonlinear refractive index which is essential for quantum technological applications. It also provides coherent control of optical properties of plasmonic nanostructures with applications in loss-compensated propagation and zero-index to high-refractive-index plasmonic metamaterials, as well as photonic switches.
\end{abstract}

\maketitle

\section{Introduction}

The ability to actively control the refractive index of materials is a cornerstone of modern photonic and optoelectronic technologies, offering enhanced performance and adaptability across a wide range of applications such as optical communications, sensing and ultrafast quantum switching \cite{Davis2014,Fang2015, Stewart2008}.
Traditionally, the refractive index of a material is considered static under fixed environmental conditions. However, achieving continuous and tunable enhancement in the refractive index opens new pathways for dynamically controlling light-matter interactions in real time, enabling new paradigms in nonlinear optics, adaptive photonic devices, and ultrafast optical switching \cite{Chai2016}.\\
One approach to achieving tunable optical properties involves the periodic arrangement of materials with high and low refractive indices to form photonic crystal structures \cite{Judith1998}. While this method enables a degree of refractive index modulation, it is often constrained to a limited range of 5–10$\%$ and typically lacks reversibility, restricting its adaptability for dynamic applications. Metamaterials, on the other hand, offer even greater flexibility by enabling the realization of exotic optical properties, such as negative refractive index \cite{Liberal2017} and hyperbolic dispersion. They have also proven crucial for enhancing nonlinearities, surface enhanced spectroscopies, and ultrafast optical switching \cite{Kuznetsov2024, Sergei2019}.\\
Recent advances in the metamaterials functional components, particularly meta-atoms \cite{Meinzer2014}, have significantly enhanced the optical properties of materials with tunable characteristics \cite{FAN2015, Soukoulis2011, Asif2023, Asif2024, Giannini2011}. For instance, plasmonic metasurfaces have been employed for the quantum optical \textit{enhancement of index of refraction} (EIR), enabling all-optical switching mechanisms \cite{Dhama2022}. This plasmonic analogue of index enhancement not only allows for amplitude modulations of optical signals but also creates a stop band at the desired frequency with a large band gap \cite{Yuce2021,Scully1991}. Recently, Panahpour et al proposed EIR through the coherent control of plasmon resonances in silver nanoantenna structures. Using a classical approach, they demonstrated linear enhancement of the refractive index by controlling the polarizability and absorption of the plasmonic nanoantennas using a probe pulse  \cite{Panahpour2019}.\\ 
However, the classical treatment of plasmonic analogue of EIR fails to account for second quantization or nonlinear effects in the system which are very crucial for quantum entanglement \cite{Mehmet2023} and harmonic generation in photonic devices. Moreover, previous studies have primarily focused on the amplitude modulation in EIR, limiting the continuous spectral tunability of photonic systems. Normally, resonant systems operate at a particular frequency (i.e. at resonance), which corresponds to the point where signal-to-noise ratio is high and the system's response is stronger. If the system is driven at a different frequency, performance declines, either due to an increase in noise, or a weakening of the system’s response. While tunability of resonance frequency has been explored in the infrared regime using graphene nanostructures \cite{Lee2013}, this approach is less applicable to photonic systems operating in the visible region.\\ 

\begin{figure}
\includegraphics[scale=0.4]{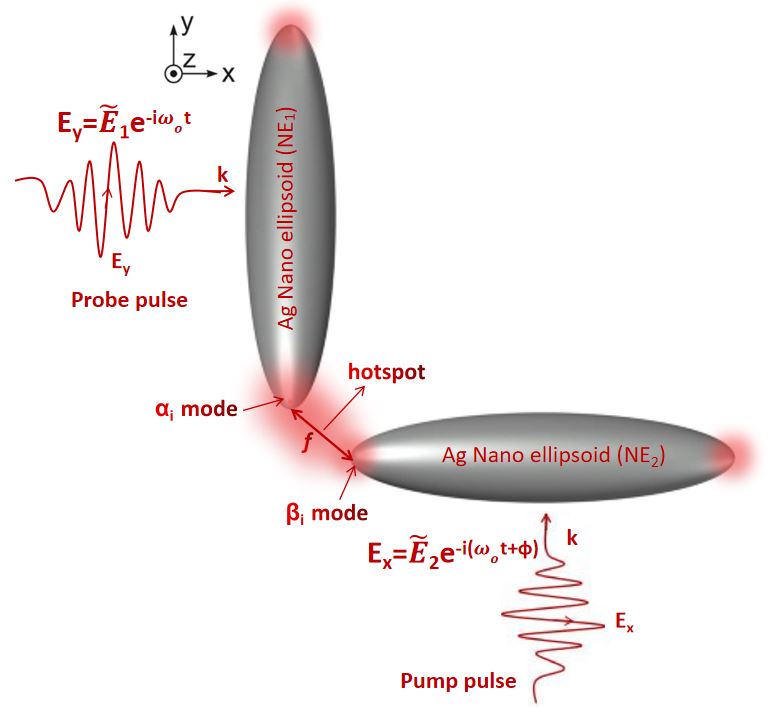}
\caption{\label{fig:1} Schematic diagram of silver nano-ellipsoids in rectangular configuration, excited through pump (x-polarized) and probe  (y-polarized) pulses. We consider spheroidal nanoparticles with a major semi-axis of 30 nm and a minor semi-axis of 10 nm.}
\end{figure}

To address all these discrepancies, here, we demonstrate coherent optical control of the EIR in both linear and nonlinear plasmonic systems through a quantum mechanical approach for the first time to the best of our knowledge. The advantage of using a quantum mechanical approach for index enhancement is that it can be applied to both linear and nonlinear systems, regardless of the system dynamics, in contrast to classical approach. Moreover, this approach allows for tuning the driving frequency of the system, enabling it to behave as though it were resonant without requiring the development of a new device. Furthermore, our proposed system operates entirely within the visible spectrum (around 500 nm) using silver nanoparticles, which is one of its most valuable aspects.
 
In the modal system, we employ two nano-ellipsoids (NEs) as plasmonic nanoantennas arranged in a rectangular configuration, as previously outlined \cite{Panahpour2019, Yuce2021, Gunay2020}. NE$_1$, aligned in the y-direction, is excited by a y-polarized probe pulse with amplitude $\tilde{E}_1$, while NE$_2$ aligned in the x-direction, is driven by an x-polarized pump pulse with amplitude $\tilde{E}_2$ and phase difference $\phi$. The schematic diagram of this coupled system is shown in Fig.\ref{fig:1}. The probe pulse excites linear plasmon mode ($\alpha_1$) along the axes of NE$_1$, while the pump pulse induces polarization in NE$_2$ with amplitude $\beta_1$. Under the linear coupling of both amplitudes ($\alpha_1$ and $\beta_1$), we aim to investigate the modulation of EIR of the probe field by tuning amplitude and phase of the control pulse. In both linear and nonlinear systems, we first drive the results through analytical approach and then verify it through finite difference time domain (FDTD)simulations.

\section{Linear enhancement of index of refraction}
In this section, we investigate the linear enhancement of the index of refraction by analyzing the steady-state response of plasmon amplitudes ($\alpha_1$,$\beta_1$) supported by NE$_1$ and NE$_2$, respectively. The dynamics of the coupled system of NEs are derived through a quantum mechanical approach. To begin, we define the total Hamiltonian of the system within the framework of quantum harmonic oscillator model, where the first order quantized modes of NEs are linearly coupled. The total Hamiltonian for the linear system ($\hat{\mathcal{H}}_{l}$) is given as follow,

\begin{eqnarray}
	\hat{\mathcal{H}}_{l}=\hbar\omega_{a}\hat{a}_{1}^\dagger\hat{a}_{1}+\hbar\omega_{b}\hat{b}_{1}^\dagger\hat{b}_{1}+ i \hbar \tilde{E}_1 e^{-i\omega_o t}\hat{a}_{1}^\dagger \nonumber \\ + i \hbar \tilde{E}_2 e^{-i(\omega_o t+\phi)}\hat{b}_{1}^\dagger+\hbar f (\hat{a}_{1}^\dagger{\textdagger}\hat{b}_{1}+\hat{b}_{1}^\dagger{\textdagger}\hat{a}_{1})
	\label{eq:1}
\end{eqnarray}

where the first and second terms represents the energies $\hbar\omega_{a}$ and $\hbar\omega_{b}$ of $\alpha_{1}$ and $\beta_{1}$ plasmon modes, respectively, along with the corresponding raising (lowering) operators $\hat{a}_{1}^\dagger(\hat{a}_{1})$, $\hat{b}_{1}^\dagger(\hat{b}_{1})$. The third and fourth terms describe the excitation of the $\hat{a}_{1}$ and $\hat{b}_{1}$ modes through y-polarized probe and x-polarized pump pulses, both at frequency $\omega_o$. Here, we choose the x-polarized pump pulse amplitude in the real plane ($\tilde{E_2}$ $= E_2$) and y-polarized probe pulse amplitude in the imaginary plane ($\tilde{E_1}$ $= i E_1$), for simplicity. This choice does not affect the overall results or the underlying physics of the system, as the absolute values of plasmon amplitudes are used in the analysis. The final term accounts for the interaction between the two linear modes with a coupling strength $f$. To derive the time-evolution of these plasmon modes, we solve the Hamiltonian defined in Eq.\ref{eq:1} using the Heisenberg equations of motion, $i\hbar\hat{\dot{a}}_{1}= [\hat{a}_{1},\hat{\mathcal{H}}_{l}]$ and $i\hbar\hat{\dot{b}}_{1}= [\hat{b}_{1},\hat{\mathcal{H}}_{l}]$. The resulting driven-dissipative dynamics are as follows:

\begin{equation}
	\dot{\alpha}_{1}= -(i\omega_a+\gamma_a)\alpha_{1}- i f \beta_{1} + \tilde{E}_1 e^{-i\omega_o t}
	\label{eq:2}
\end{equation}

\begin{equation}
	\dot{\beta}_{1}= -(i\omega_b+\gamma_b)\beta_{1}- i f \alpha_{1} + \tilde{E}_2 e^{-i\omega_o t-i\phi}
	\label{eq:3}
\end{equation}
where we replace the operators $\hat{a}_1$ and $\hat{b}_1 $ with the complex amplitudes $\alpha_1$ and $\beta_1 $, respectively. The decay rates $\gamma_{a}$ and $\gamma_{b}$ corresponds to the damping of the $\alpha_1$ and $\beta_1 $ modes, respectively. In the steady-state, by employing $\alpha_1(t)= \alpha_1 e^{-i\omega_o t}$ and $\beta_1(t)=\beta_1 e^{-i\omega_o t}$ the complex amplitudes of both modes takes the following form: 

\begin{equation}
	\alpha_1= \frac{-i f \tilde{E}_2 e^{-i\phi} + \tilde{E}_1 \delta_2}{\delta_1 \delta_2 + |f|^2}, 
	\label{eq:4}
\end{equation}
\begin{equation}
	\beta_1= \frac{-i f^* \tilde{E}_1 e^{-i\phi} + \tilde{E}_2 \delta_1 e^{-i\phi} }{\delta_1 \delta_2 + |f|^2},
	\label{eq:5}
\end{equation}

Here, $\delta_1 = [i (\omega_a-\omega_o)+\gamma_a]$ and $\delta_2 = [i (\omega_b-\omega_o)+\gamma_b]$ represent the detuning parameters. For simplicity, all parameters, including decay rates ($\gamma_{a,b}=0.05\omega_o$) and plasmon resonance frequencies ($\omega_a=\omega_b=\omega_o$), are scaled with respect to the resonance frequency of the dimer, $\omega_o = 3.15\times 10^{15}$ rad/sec. The coupling strength $f$ is taken as $0.06\omega_o$ \cite{Panahpour2019} and the amplitude of probe pulse E$_1$ is set to $10^7$ V/m. The complex plasmon amplitudes reflect the strength of polarization in NEs. To show the enhancement in the refractive index ($n$), we analytically evaluate its real $(n)$ and imaginary $(\kappa)$ parts using Eqs.\ref{eq:6} and \ref{eq:7} \cite{Fleischhauer1992}. These results are then compared with the numerical results obtained through finite difference time domain (FDTD) simulations. In the simulation, we evaluate the polarization by normalizing it to source position ($e^{ikz}$). The position of the source does not impact the enhancement or spectral shape in the polarized field therefore we measure the average value of field intensity by placing point monitors at different positions around the nanoparticle to reduce the total simulation time.

\begin{equation}
n_{real}\propto Re(\sqrt{\alpha_i/E_i}),  \hspace{0.5cm} n_{imag}\propto Im(\sqrt{\alpha_i/E_i}),   
\label{eq:6}
\end{equation}

\begin{equation}
n_{real}\propto Re(\sqrt{\beta_i/E_i}),  \hspace{0.5cm} n_{imag}\propto Im(\sqrt{\beta_i/E_i}),   
\label{eq:7}
\end{equation}

\begin{figure}
    \centering
    \includegraphics[width=1\linewidth]{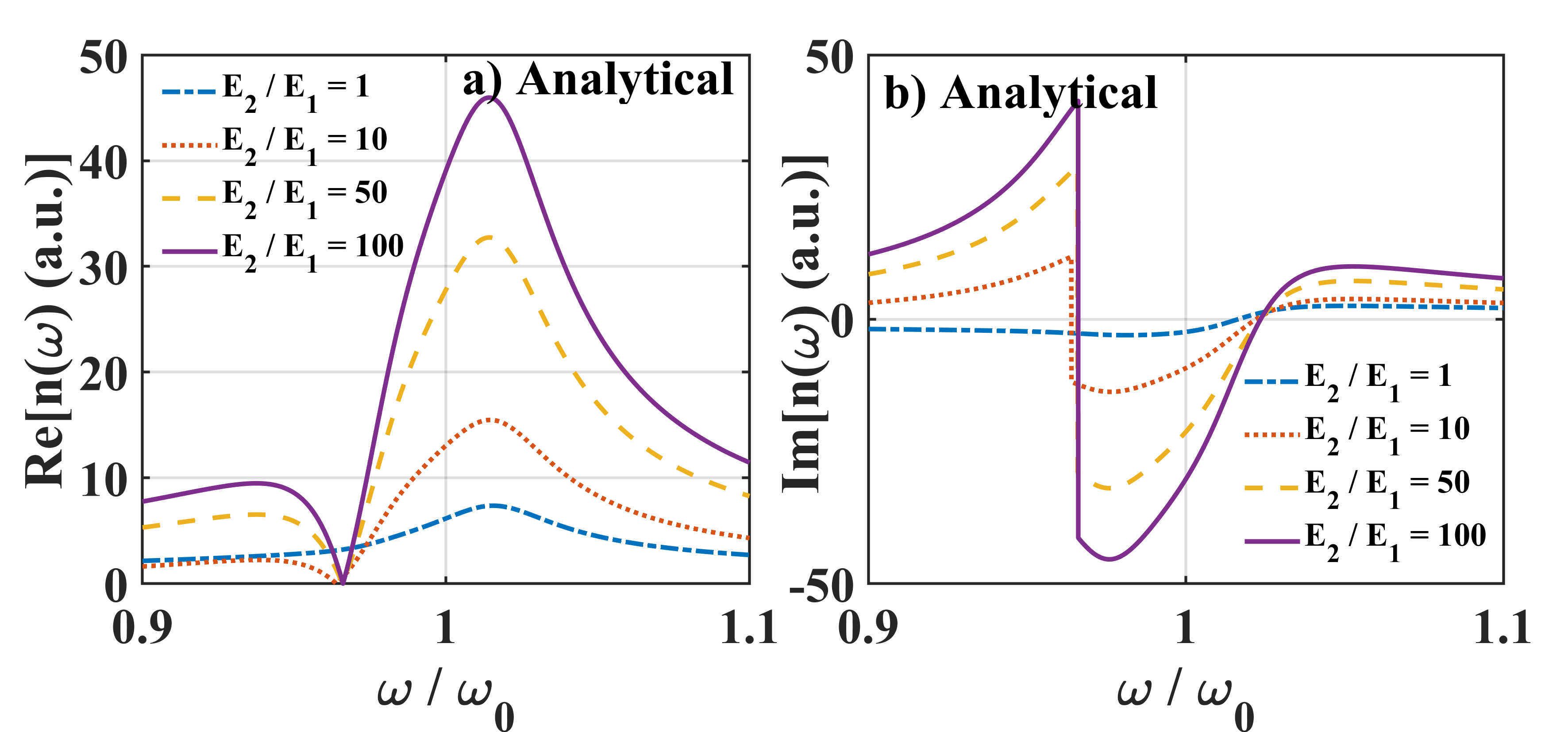}
   
   \includegraphics[width=1\linewidth]{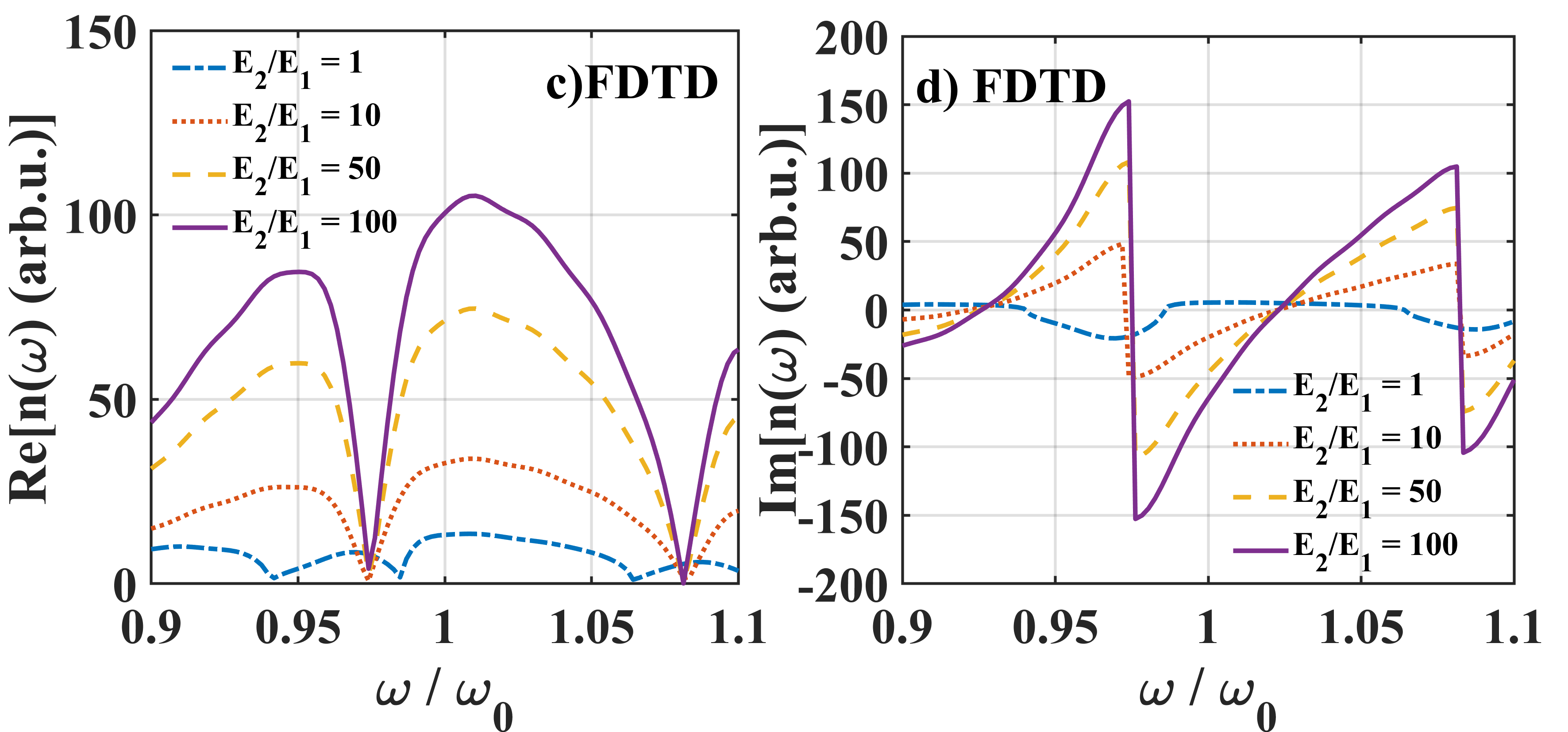}
    
        \caption{\label{2.2.png}Enhancement in the real and imaginary parts of refractive index ($n$) in the linear regime as a function of normalized frequency for different pulse amplitudes, The results are obtained using (a,b) analytical method and (c,d) FDTD simulations. $\phi$ is taken as $\pi/2$.}
    \
\end{figure}

Here $i=1,2$ denote the indices corresponding to the linear and nonlinear components of the plasmon amplitudes. We plot the real (Re[$n(\omega)]$) and imaginary (Im[$n(\omega)]$) parts of the refractive index as a function of the normalized frequency, $\omega/\omega_{o}$, for different ratios of pulse amplitudes (E$_2$/E$_1$), as shown in Fig.\ref{2.2.png}(a-d). As illustrated in Fig.\ref{2.2.png}(a), an increase in the probe pulse amplitude leads to an enhancement in the real part of $n$, which is directly related to the strength of polarization in the nano-ellipsoids. This enhanced polarization strength correlates with an increase in the index of refraction, thereby validating the classical counterpart of EIR \cite{Panahpour2019}. Notably, when the amplitude of the $x$-polarized control pulse is varied, significant changes are observed in the system without modifying any parameter of the source. This variation indicates ability of the control source, which under normal conditions would not effect the probe's polarization, to induce a change in the probe’s behavior.

\begin{figure}
    
		\centering
    \includegraphics[width=1\linewidth]{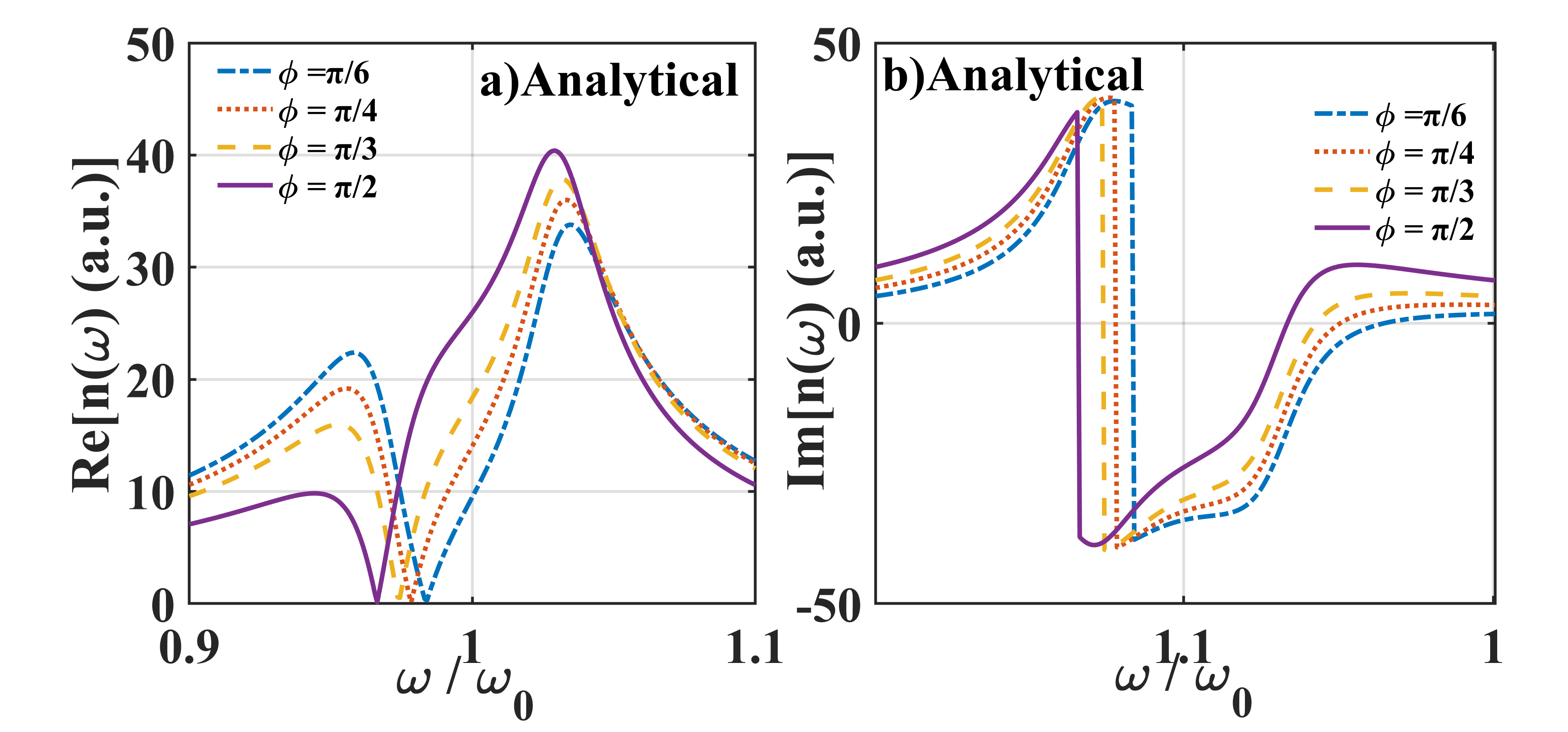}
\includegraphics[width=1\linewidth]{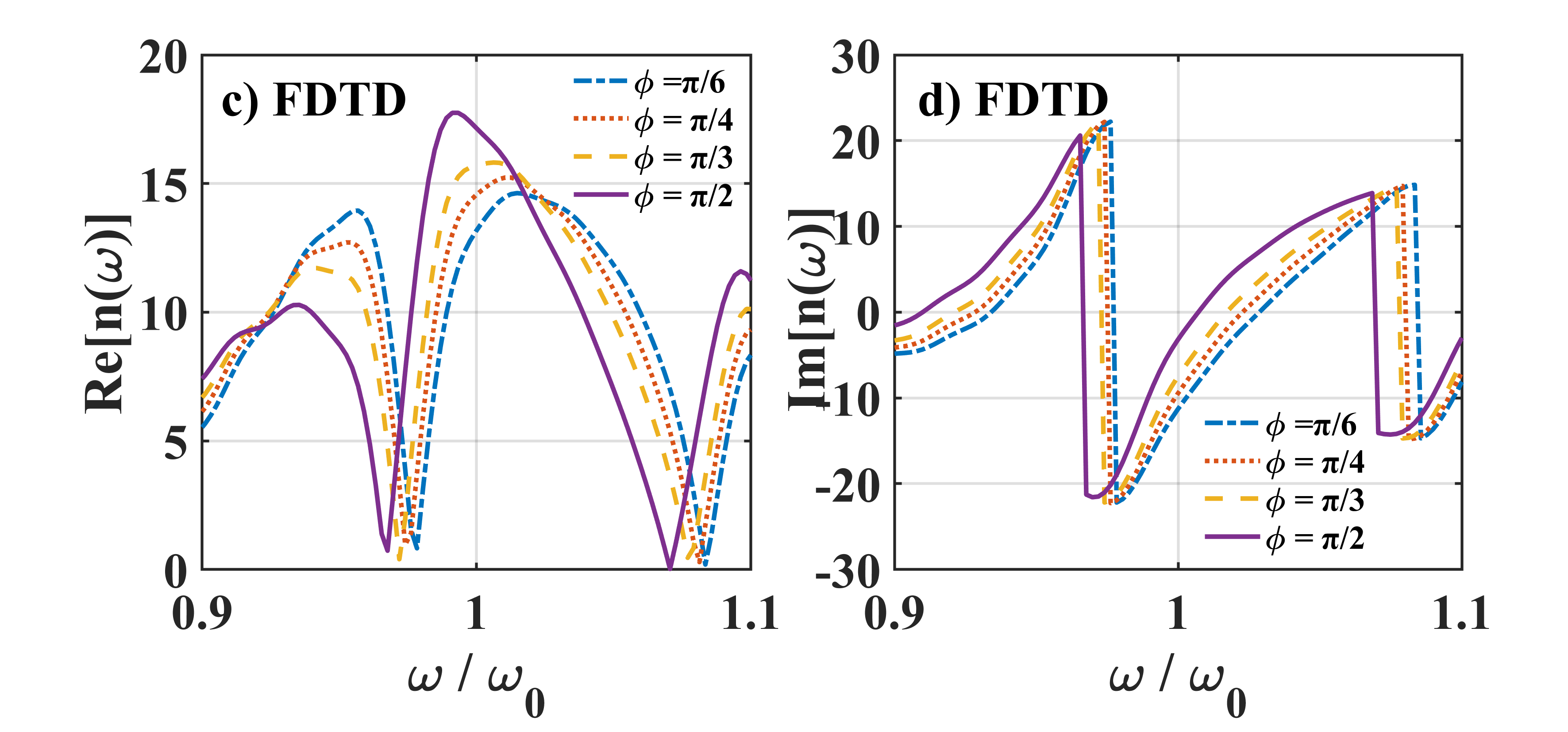}
            \caption{\label{3.2.png}Enhancement in the real and imaginary parts of the refractive index ($n$) in the linear regime as a function of normalized frequency for various phase differences ($\phi$) between probe and pump pulse. Results are presented using (a,b) analytical approach and (c,d) FDTD simulations. The pulse amplitude ratio is E$_1=50$E$_2$.}
    \

\end{figure}

In the proposed system, the refractive index is enhanced approximately an order of magnitude through amplitude modulation, achieved solely by illuminating the system with a coherent source. This enhancement is attributed to the indirect effect of changes at the hotspot on the probe, even though the control source does not directly contribute to the polarization at the ends of the long axis of the y-axis rod. In Fig.\ref{2.2.png}(a), a six-fold enhancement in the refractive index occur for probe amplitude E$_2$=100E$_1$ at frequency $\omega=1.01\omega_o$ in contrast to Re$[n(\omega)]$ for E$_2$=E$_1$. Similarly, the FDTD simulation result (Fig.\ref{2.2.png}(c)) shows an 8-fold enhancement in Re$[n(\omega)]$ for the maximum amplitude of the probe pulse at the same frequency. On the other hand, the imaginary part of n becomes negative from $0.97\omega_o$ to $1.01\omega_o$ for E$_2$=100E$_1$ indicating zero absorption of the probe pulse in the system yielding to index enhancement [see Fig.\ref{2.2.png}(b,d)]. 
 
To analyze the impact of phase modulation on the index enhancement, we evaluate the real and imaginary parts of $n$ for different phases of the probe pulse using both analytical and numerical approaches. The results are plotted in Fig.\ref{3.2.png}(a-d). As the phase varies from $\pi/6$ to $\pi/2$, analytical and FDTD results show an order of magnitude enhancement in Re$[n(\omega)]$ at frequency $0.98\omega_o$ [see Fig.\ref{3.2.png}(c)] along with spectral shifts in the peak positions. These spectral shifts are also observed in the imaginary counterparts of n for different values of probe phase, as shown in Fig.\ref{3.2.png}(b,d).    

\begin{figure}
    \centering
    \includegraphics[width=1\linewidth]{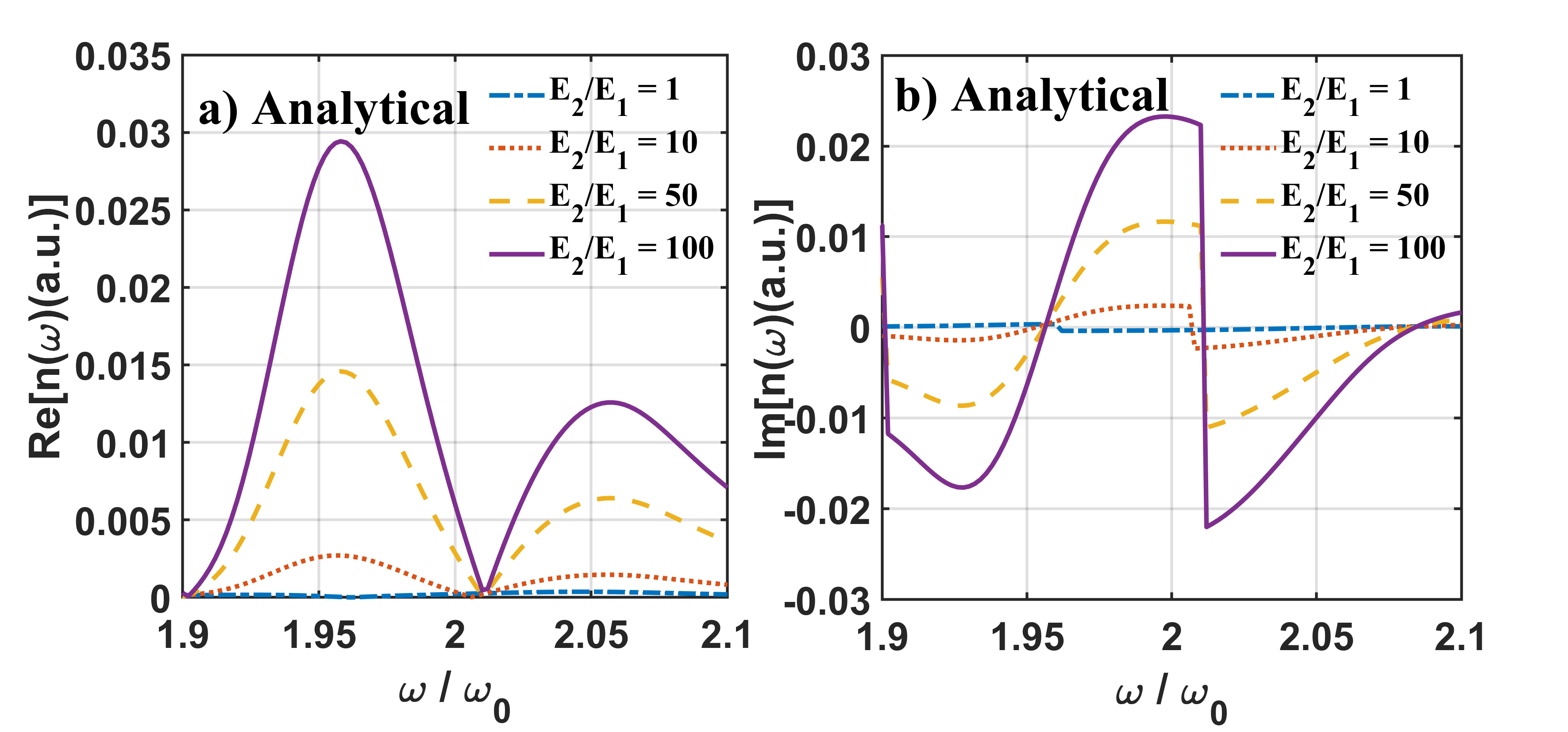}
       \includegraphics[width=1\linewidth]{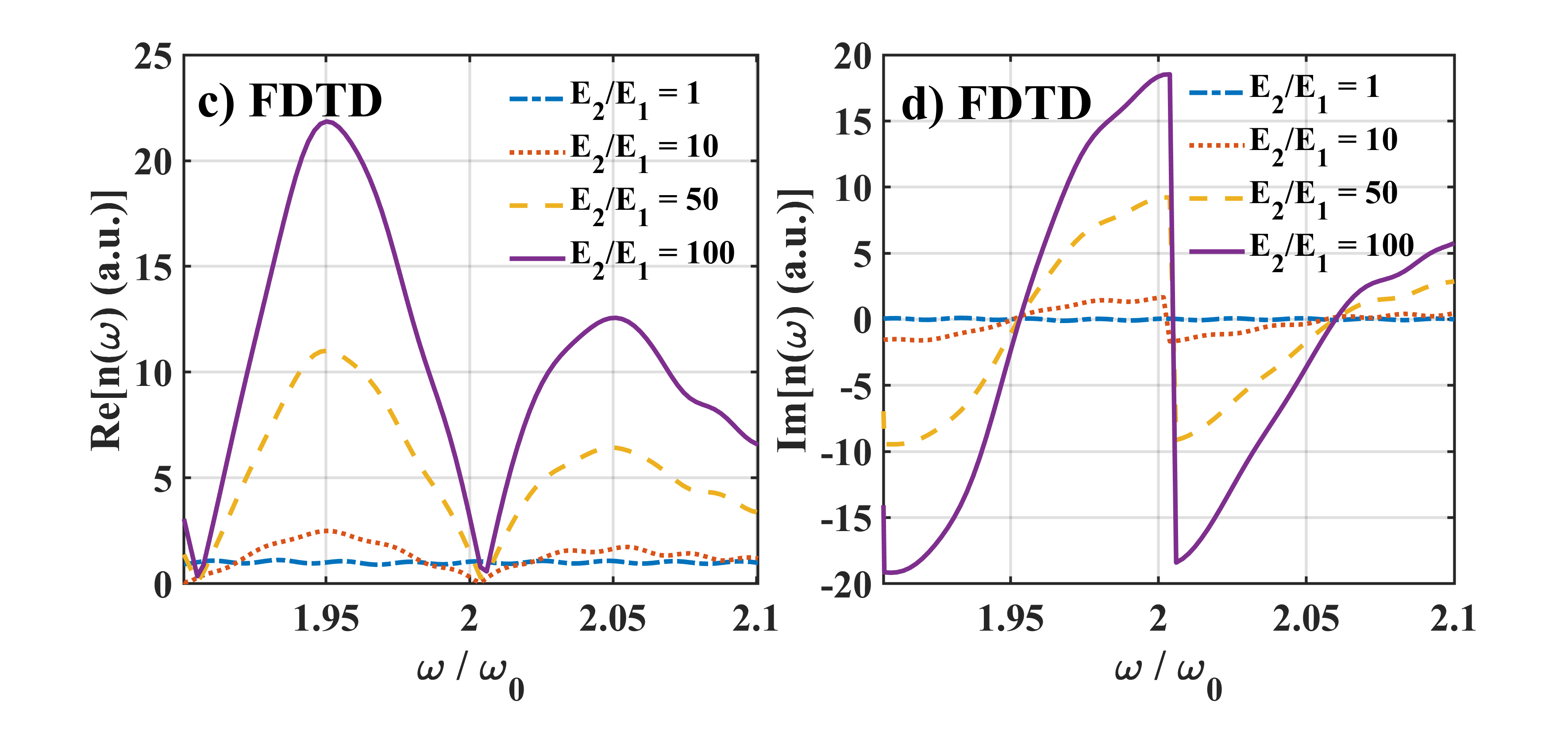}            
			\caption{\label{4.2.png}Enhancement in the real and imaginary parts of refractive index ($n$) in the non-linear regime as a function of normalized frequency for different ratios of pulse amplitudes. Results are evaluated  through (a,b) analytical method and (c,d) FDTD simulations. $\phi$ is set to $\pi/2$.}
\end{figure}
    
\section{Nonlinear enhancement of index of refraction}
In this section, we investigate a nonlinear system exhibiting EIR and investigate the tunability of its nonlinear modes indirectly. For the second-order quantization, we consider a system of two coupled NEs that support nonlinear plasmon modes. A nonlinear mode in NEs arises due to the overlap of an intense linear plasmon modes excited by both the pump and probe pulses \cite{Fanoresonance2020}. The schematic of this system remain similar to the one shown in the previous section, with the inclusion of nonlinearity [see Fig.\ref{fig:1}]. In this framework, we define the total Hamiltonian ($\hat{\mathcal{H}}_{t}$) that in-cooperates the quantized energies of both linear and nonlinear plasmonic modes supported by the NEs. The $\hat{\mathcal{H}}_{t}$ is expressed as follow,

\begin{eqnarray}
\hat{\mathcal{H}}_{t}= \hat{\mathcal{H}}_{l} + \hbar\Omega_{a}\hat{a}_{2}^\dagger\hat{a}_{2}+\hbar\Omega_{b}\hat{b}_{2}^\dagger\hat{b}_{2} + \hbar g (\hat{b}_{2}^\dagger{\textdagger}\hat{a}_{2}+\hat{a}_{2}^\dagger{\textdagger}\hat{b}_{2})  \nonumber\\ + \hbar \chi^{(2)} (\hat{a}_{1}^\dagger{\textdagger}\hat{a}_{1}^\dagger{\textdagger}\hat{a}_{2}+\hat{a}_{2}^\dagger{\textdagger}\hat{a}_{1}\hat{a}_{1}) + \hbar \chi^{(2)} (\hat{b}_{1}^\dagger{\textdagger}\hat{b}_{1}^\dagger{\textdagger}\hat{b}_{2}+\hat{b}_{2}^\dagger{\textdagger}\hat{b}_{1}\hat{b}_{1})
\label{eq:8}
\end{eqnarray}

\begin{figure}

    \centering
    \includegraphics[width=1\linewidth]{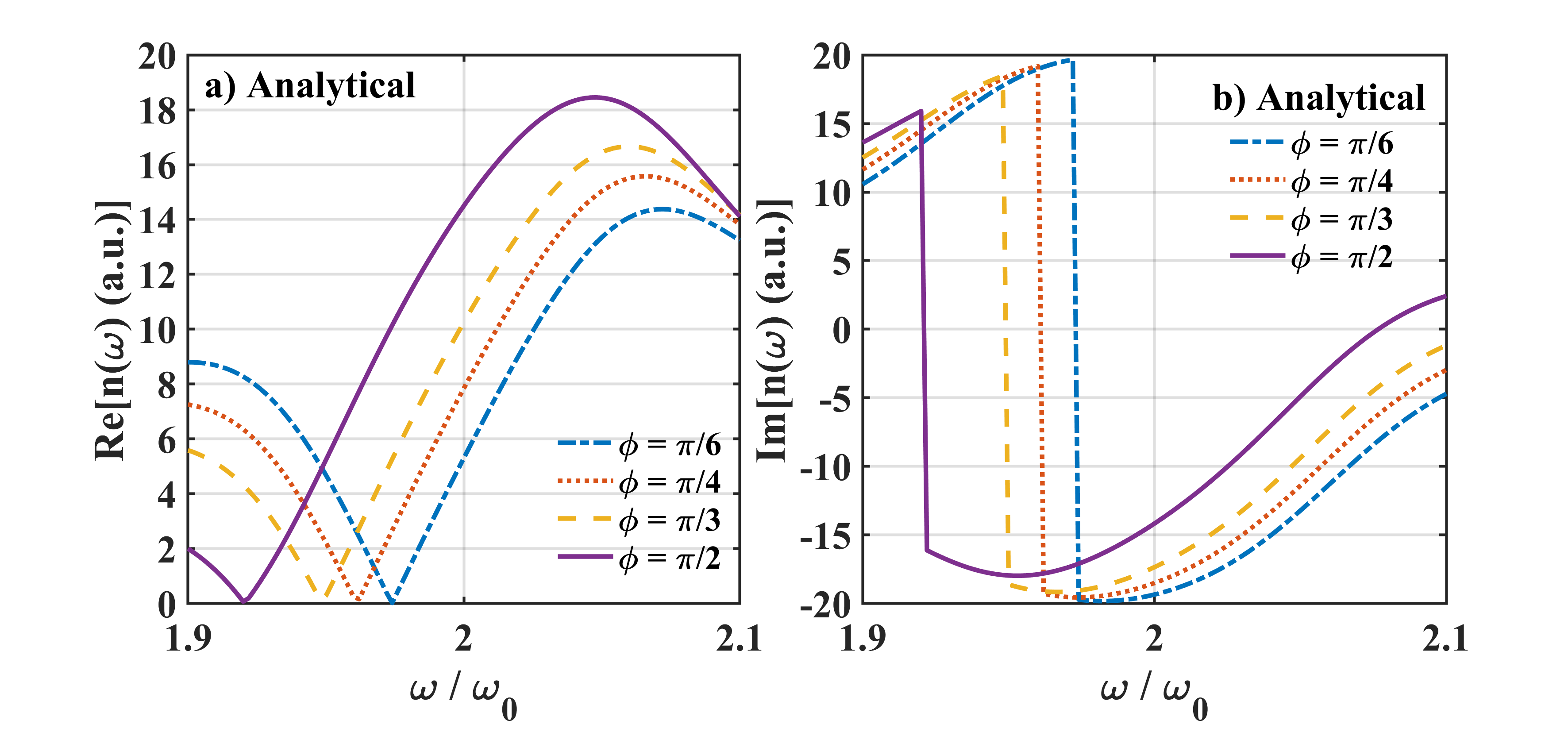}
    \includegraphics[width=1\linewidth]{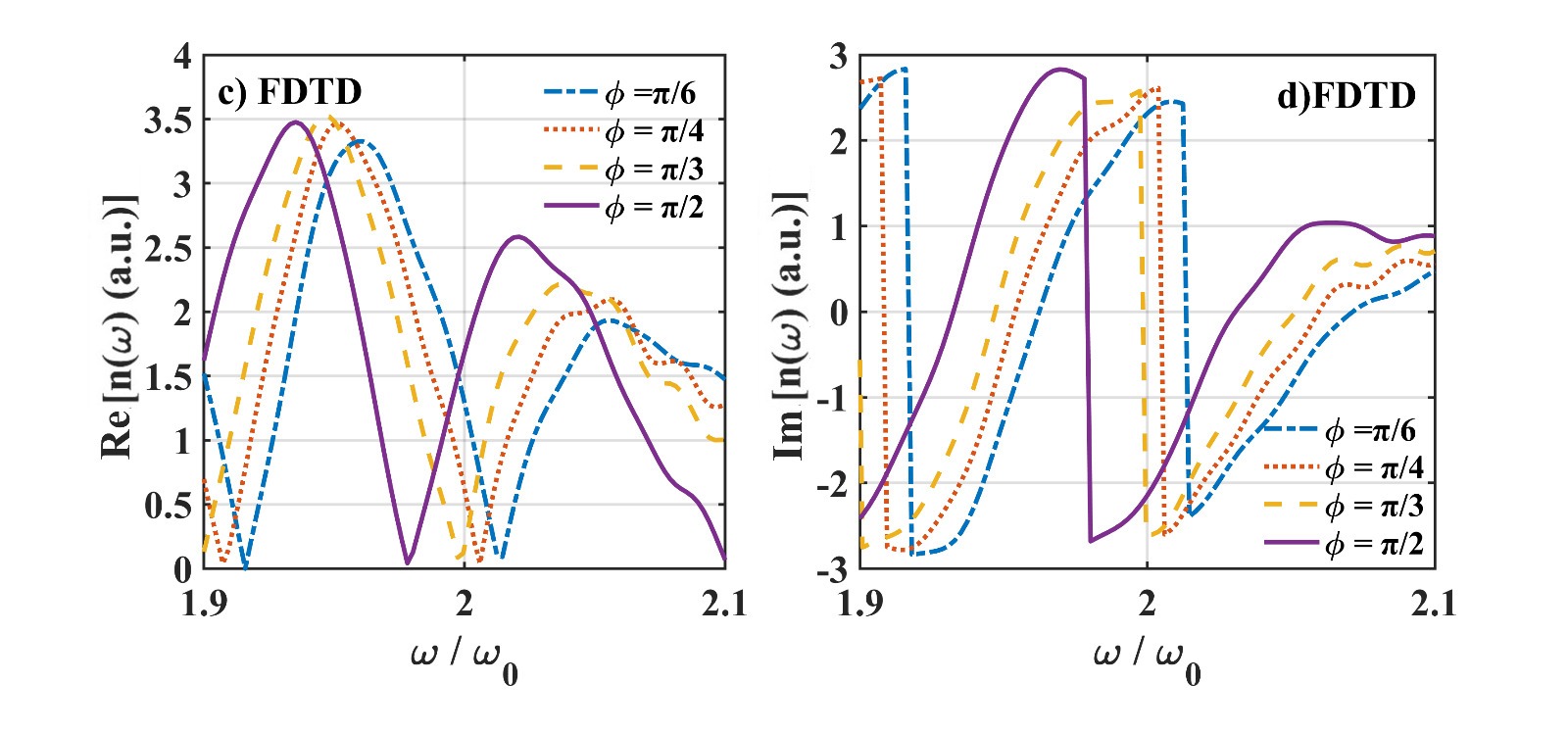}
      
\caption{\label{5.2.png} Enhancement in the real and imaginary parts of refractive index ($n$) in the non-linear regime as a function of normalized frequency for different phase difference of the probe pulse. Panels (a, b) show results obtained using analytical methods, while (c,d) present results from FDTD simulations. The field ratio E$_2/$E$_1$ is taken as 50.}
\end{figure}

Here, $\hat{\mathcal{H}}_{l}$ represents the Hamiltonian of the linear system, as defined in Eq.\ref{eq:1}. The terms $\hbar\Omega_a$ and $\hbar\Omega_b$ denote the energies of the second-harmonic plasmon modes supported by NE$_1$ and NE$_2$, respectively, with $\hat{a}_{2}^\dagger, \hat{b}_{2}^\dagger$ as the creation and $\hat{a}_{2}$,$\hat{b}_{2}$ as the annihilation operators, respectively for the nonlinear plasmon modes. The parameter g is the coupling constant that quantifies the interaction strength between the nonlinear modes of the two NEs. In this analysis, we assume that the nonlinear plasmon fields do not interact with each other, i.e., g$=0$ since these modes are not as strong as the fundamental plasmon modes. The last two terms in the Hamiltonian describe the hybridization of linear modes ($\hat{a}_{1}$,$\hat{b}_{1}$) to generate nonlinear modes ($\hat{a}_{2}$,$\hat{b}_{2}$) through a second-order nonlinear susceptibility tensor $\chi^{(2)}$ which governs the strength of the nonlinear signal derived from the overlap integral \cite{Asif2022}. We solve the Hamiltonian using the Heisenberg equations of motion, $i\hbar\hat{\dot{a}}_{j}= [\hat{a}_{j},\hat{\mathcal{H}}_{t}]$ and $i\hbar\hat{\dot{b}}_{j}= [\hat{b}_{j},\hat{\mathcal{H}}_{t}]$, and drive the time-evolution of the linear and nonlinear plasmon modes. Here, the operators are replaced with complex amplitudes $\alpha_i$ and $\beta_i$, (where $i=1,2$), as shown below,

\begin{equation}
 \dot{\alpha}_{1}= -(i\omega_a+\gamma_a)\alpha_{1}- i f \beta_{l} +  E_1 e^{-i\omega t}- i \chi^{(2)} \alpha_{1}^* \alpha_{2}
 \label{eq:9}
\end{equation}
\begin{equation}
 \dot{\beta}_{l}= -(i\omega_b+\gamma_b)\beta_{l}- i f \alpha_{1} + E_2 e^{-i\omega t-i\phi} - i \chi^{(2)} \beta_{1}^* \beta_{2}
 \label{eq:10}
\end{equation}
\begin{equation}
 \dot{\alpha}_{2}= -(i\Omega_a+\Gamma_a)\alpha_{2}- i g \beta_{2} - i \chi^{(2)} \alpha_{1}^2 
 \label{eq:11}
\end{equation}
\begin{equation}
 \dot{\beta}_{2}= -(i\Omega_b+\Gamma_b)\beta_{2}- i g \alpha_{2} - i \chi^{(2)} \beta_{1}^2 
 \label{eq:12}
\end{equation}

where $\Gamma_a$ and $\Gamma_b$ represent the decay rates of the nonlinear plasmon modes. The complex amplitudes of plasmons modes are derived by solving the time-dependent differential equations using Runge-Kutta method in MATLAB. The frequency and decay rate parameters are set as $\Omega_a=\Omega_b=2\omega_o$ and $\Gamma_{a,b}=\gamma_{a,b}=0.06\omega_o$ with $f = 0.05\omega_o$ \cite{Panahpour2019}. The nonlinear susceptibility is taken as $\chi^{(2)}=10^{-10}\omega_o$. To analyze the changes in the nonlinear EIR, we evaluate the real and imaginary parts of refractive index using Eq.\ref{eq:6} and Eq.\ref{eq:7} and plot the results, as shown in Fig.\ref{4.2.png} and Fig.\ref{5.2.png}.
For different amplitudes, the curves exhibit similar trends in both approaches. However, the results obtained through the FDTD simulations demonstrate a significant enhancement in the real and imaginary parts of refractive index at second-order quantization. The difference in amplitudes in analytical and FDTD results is due to higher amplitude of probe pulse used in the simulations.

In Fig.\ref{4.2.png}, a significant modulation in the index is observed for relatively large values of amplitude and phase of probe pulse, while the results remain closely matched for small variations.  The nonlinear plasmon amplitude ($\alpha_2/\beta_2$) is significantly smaller (not shown here) compared to the fundamental plasmon amplitude ($\alpha_1/\beta_1)$. This suggests that nonlinear interactions can be neglected at small values of (E$_2/$E$_1$). For E$_2$=E$_1$, more than an order of magnitude enhancement occur in Re$[n(2\omega)]$ at frequency $1.95\omega_o$ for both analytical and FDTD results, as shown in Fig.\ref{4.2.png}(a),(c).
Whereas, the Im$[n(2\omega)]$ part becomes zero or negative at the position of maximum $n$ [see Figures.\ref{4.2.png}(b,d)]. In Fig.\ref{5.2.png}(a,c), variations in the phase difference not only enhance the refractive index but also induce a significant spectral shift in the system's operational frequency for large phase differences. At the second harmonic frequency ($\omega=2\omega_o$), the index experienced by the pump enhances up to 3-fold while variation in the probe phase from $\pi/2$ to $\pi/6$ yields a redshift in spectral position. This spectral shift driven by phase modulation, highlights the potential for on-demand, continuous tunability of the system's optical properties, making it highly compatible for integrated photonic applications.

Our model demonstrates a proof of principle concept for on-demand coherent quantum control of optical systems operating at visible frequencies. Beyond providing a controllable mechanism for refractive index modulation, the ability to dynamically tune the system's resonant frequency offers new opportunities for applications in devices such an photonic integrated circuits (PICs), metasurfaces and quantum computing systems. Furthermore, the significant nonlinear index enhancement and spectral tunability observed in our model system provides a powerful avenue for active tuning of Epsilon-Near-Zero (ENZ) materials. This capability allows precise control over light-matter interactions, enabling dynamical modulation of optical properties critical for reconfigurable photonic devices and next-generation optical systems.    

\section{Conclusion} 
We demonstrated active optical control of index enhancement using a quantum mechanical approach. By modulating the phase and amplitude of the pump pulse, we analyzed both linear and nonlinear enhancements in the refractive index. For varying pulse amplitudes, maximum enhancement in index of refraction is observed in both linear and nonlinear regimes. In addition to index enhancement, variation in pulse phases cause a blueshift in the peak spectral positions across multiple orders, highlighting the system's continuous tunability in resonance frequency of the system in the visible regime. Our setup not only enables precise refractive index modulation but also offers dynamic control over the system's operation frequency even in the nonlinear regime. This dual capability paves the way for innovative applications across a range of technologies, including photonic devices, integrated circuits, and quantum systems.

\begin{acknowledgments}
R.S., M.E.T., M.G., E.O., and H.A. acknowledge support from TUBITAK No. 123F156.
\end{acknowledgments}
\normalsize $^{\dagger}$These authors equally contributed to this work. 

\bibliography{EIR}

\end{document}